\documentclass[a4paper,10pt,showpacs,notitlepage,nofootinbib,floatfix,superscriptaddress,prd]{revtex4-1}
\pdfoutput=1
\usepackage{amsfonts}
\usepackage{amsmath}
\usepackage{graphicx}
\usepackage{dcolumn}
\usepackage{bm}
\usepackage{amssymb}%
\usepackage{slashed}
\usepackage{color}

\def\beq{\begin{equation}}
\def\eeq{\end{equation}}

\def\bea{\arraycolsep .1em \begin{eqnarray}}
\def\eea{\end{eqnarray}}

\def\eps{\epsilon}

\def\al#1{\alpha_{#1}}
\def\eq#1{(\ref{#1})}

\def\s0#1#2{\mbox{\small{$ \frac{#1}{#2} $}}}
\def\0#1#2{\frac{#1}{#2}}

\def\grgl{\:\hbox to -0.2pt{\lower2.5pt\hbox{$\sim$}\hss}{\raise3pt\hbox{$>$}}\:}
\def\klgl{\:\hbox to -0.2pt{\lower2.5pt\hbox{$\sim$}\hss}{\raise3pt\hbox{$<$}}\:}

\def\lsim{\mathrel{\rlap{\lower4pt\hbox{\hskip1pt$\sim$}}
    \raise1pt\hbox{$<$}}}                
\def\gsim{\mathrel{\rlap{\lower4pt\hbox{\hskip1pt$\sim$}}
    \raise1pt\hbox{$>$}}}                

\newcommand{\DDslash}[1]{\hspace*{0.1cm} \slash \hspace*{-0.26cm}{#1} }
\begin{document}
\title{Thermodynamics of asymptotically safe theories}
\author{Dirk H.\ Rischke}
\affiliation{Institute for Theoretical Physics, Goethe University,
Max-von-Laue-Str.\ 1, D--60438 Frankfurt am Main, Germany }
\author{Francesco Sannino}
\affiliation{CP$^3$-Origins \& the Danish Institute for Advanced Study, Danish IAS, 
University of Southern Denmark, Campusvej 55, DK--5230 Odense, Denmark}

\begin{abstract}
We investigate the thermodynamic properties of a novel class of gauge-Yukawa theories 
that have recently been shown to be completely asymptotically safe, because their 
short-distance behaviour is determined by the presence of an interacting fixed point.  
Not only do all the coupling constants freeze at a constant and 
calculable value in the ultraviolet, their values can even
be made arbitrarily small 
for an appropriate choice of the ratio $N_c/N_f$ of fermion colours and flavours in the 
Veneziano limit. Thus, a perturbative treatment can be justified. 
We compute the pressure, entropy density, and thermal degrees of freedom
of these theories to next-to-next-to-leading order in the coupling constants.  \\
[.1cm]
{\footnotesize  \it Preprint: CP$^3$-Origins-2015-19 DNRF90 \& DIAS-2015-19 }
\end{abstract}

\maketitle

\section{Introduction}

Theories featuring gauge bosons, fermions, and scalars constitute the backbone 
of the standard model of particle interactions. This is to date one of the most 
successful models of nature. The recent discovery of asymptotically safe quantum field
theories in four space-time dimensions \cite{Litim:2014uca}, including their
quantum-corrected potentials \cite{Litim:2015iea}, widens the horizon of fundamental theories
that can be used beyond the traditional asymptotically free paradigm 
\cite{Gross:1973id,Politzer:1973fx}. The novelty resides in the occurrence of an exact
interacting ultraviolet (UV) fixed point rather than a UV non-interacting fixed point, as it is the 
case for asymptotically free theories. 

For the class of theories we will be investigating here
{a crucial property was unveiled in 
Ref.\ \cite{Litim:2014uca}}: the Yukawa interactions, mediated by the scalars, compensate for 
the loss of asymptotic freedom due to the large number of gauged fermion flavours and 
therefore cure the subsequent growth of the gauge coupling. 
The further interplay of the gauge, Yukawa, and scalar interactions ensures that all 
couplings reach a stable interacting UV fixed point allowing for a 
{\it complete asymptotic safety} scenario in all couplings \cite{Litim:2014uca}. 
This is different from the {\it complete asymptotic freedom} scenario 
\cite{Gross:1973ju,Cheng:1973nv,Callaway:1988ya} where all couplings 
vanish in the UV, see Refs.\ \cite{Holdom:2014hla,Giudice:2014tma} 
for recent studies. 

The phase diagram including the scaling exponents of the theory was 
determined to the maximum known order in perturbation theory 
\cite{Litim:2014uca,Litim:2015iea}. It was also shown in Ref.\ \cite{Litim:2015iea}  
that the scalar potential is stable at the classical and quantum level.  
Therefore these theories hold a special status: at arbitrarily short scales 
and without assuming additional symmetries they are fundamental according to 
Wilson's definition.

Having at our disposal new classes of fundamental theories one can use them to construct 
new dark-matter paradigms \cite{Sannino:2014lxa} or even support cosmic inflation 
\cite{Nielsen:2015una}.    
It is therefore timely as well as theoretically and phenomenologically relevant to investigate, in 
a controllable manner, the thermodynamics of four-dimensional completely asymptotically safe 
theories\footnote{The thermodynamics of asymptotically free theories featuring perturbatively 
controllable and interacting IR fixed points has been investigated in Ref.\ \cite{Mojaza:2010cm} 
to the maximum known order in perturbation theory.}. 

We organise this paper as follows. The theory and its salient 
zero-temperature properties are 
reviewed in Sec.\ \ref{The theory}. This is followed by the determination of the asymptotically 
safe pressure to the leading order (LO), next-to-leading (NLO), and next-to-next-to-leading 
order (NNLO) in Sec.\ \ref{AST}. The entropy density of the system is 
determined and its properties are discussed in Sec.\ \ref{entropy}.  The thermal degrees 
of freedom count for asymptotically safe theories is introduced and discussed in 
Sec.\ \ref{TDoF}. We offer our 
conclusions in Sec.\ \ref{conclusions} where we also briefly discuss the impact of introducing
a quark chemical potential. In the appendix we report the beta functions of the 
theory and further details of the computations of the pressure at the 
respective orders in perturbation theory.

\section{Zero-temperature physics}\label{SecPD}
\label{The theory}

Here we briefly review the salient aspects of the gauge-Yukawa system introduced in 
Ref.\ \cite{Litim:2014uca} such as the phase diagram of the theory and the expressions for 
the UV-safe trajectories away from the UV-stable fixed point. We will also provide the 
expressions for the running of the couplings along the globally defined UV-IR
connecting line 
known as separatrix, or line of physics. Further quantities required for the subsequent 
thermodynamical analysis will also be reported.

The asymptotically safe theory suggested in Ref.\ \cite{Litim:2014uca}
contains $N_c^2-1$ non-Abelian gauge fields, $A_\mu^i$, $N_c N_f$ 
massless Dirac fermions, $\psi$, and $N_f^2$ massless complex-valued scalars, $H$.
The theory has a local $SU(N_c)$ gauge symmetry and a global
$U(N_f)_L \times U(N_f)_R$ chiral symmetry at the classical level.  Because of the 
Adler-Bell-Jackiw axial anomaly the quantum global symmetry is 
$SU(N_f)_L \times SU(N_f)_R \times U(1)_V$. The left- and right-handed 
fermions live in the fundamental $(N_f,0)$ and $(0,N_f)$ representations of this 
symmetry group,
\begin{equation}
\psi_L \longrightarrow \psi_L^\prime = U_L \psi_L\;,\;\;\;
\psi_R \longrightarrow \psi_R^\prime = U_R \psi_R\;,
\end{equation}
while the scalars live in the adjoint $(N_f,N_f^*)$ representation,
\begin{equation}
H \longrightarrow H^\prime = U_L H \, U_R^\dagger\;.
\end{equation}
It is convenient to decompose the complex-valued $(N_f \times N_f)$ matrix $H$ in terms
of the generators $T_a$ in the fundamental representation of $U(N_f)$, $a=0, \ldots, N_f^2-1$,
\begin{equation} \label{H}
H = (S_a + i P_a) T_a\;,
\end{equation}
where $S_a$ are $N_f^2$ scalar and $P_a$ are $N_f^2$ pseudoscalar fields.
The Lagrangian reads
\begin{equation} \label{Lag}
{\cal L} =  - \frac{1}{4}\, F_{\mu \nu}^i F^{\mu \nu}_i + \bar{\psi}\, i \DDslash{D} \, \psi
+ {\rm Tr} \left( \partial_\mu H^\dagger\, \partial^\mu H \right)  
+ y \left( \bar{\psi}_L \, H \, \psi_R + \bar{\psi}_R \, H^\dagger \psi_L \right)
- u \, {\rm Tr} \left( H^\dagger H \right)^2 - v \left[ {\rm Tr} \left( H^\dagger H \right) \right]^2\;,
\end{equation}
with the covariant derivative
\begin{equation}
D_\mu = \partial_\mu + ig A_\mu^i t_i\;.
\end{equation}
Here, $g$ is the coupling constant of the non-Abelian gauge sector and $t_i, \, i = 1, \ldots,
N_c^2-1$, are the generators of $SU(N_c)$ in the fundamental representation.

At the classical level the theory counts four marginal couplings, the gauge coupling $g$, the 
Yukawa coupling $y$, the quartic scalar coupling ${u}$, 
and the double-trace scalar coupling 
$v$, which we write as:
\beq\label{couplings}
\al g=\frac{g^2\,N_c}{(4\pi)^2}\,,\quad
\al y=\frac{y^{2}\,N_c}{(4\pi)^2}\,,\quad
\al h=\frac{{u}\,N_f}{(4\pi)^2}\,,\quad
\al v=\frac{{v}\,N^2_f}{(4\pi)^2}\,.
\eeq
The appropriate powers of $N_c$ and $N_f$ in the normalization of the couplings allow to take 
the Veneziano limit of the theory.  Following Ref.\ \cite{Litim:2014uca} we will also use the 
short-hand notation $\beta_i\equiv\partial_t\alpha_i$, with $i=(g,y,h,v)$, for the beta functions 
of the respective couplings \eq{couplings}. It is convenient to introduce the continuous real 
parameter  
\begin{equation}\label{eps}
\epsilon=\frac{N_f}{N_c}-\frac{11}{2}
\end{equation}
in the Veneziano limit of large $N_f$ and $N_c$, with the ratio {$N_f/N_c$} 
fixed. The relevant 
beta functions of the theory have been obtained in Ref.\ \cite{Antipin:2013pya} in dimensional 
regularisation, using the results of 
Refs.\ \cite{Machacek:1983tz,Machacek:1983fi,Machacek:1984zw}, and 
are summarised in App.\ \ref{beta} in the Veneziano limit. 
 
For $\eps<0$ the theory is asymptotically free in the gauge sector while for 
$\eps>0$ it becomes a non-Abelian QED-like 
theory because asymptotic freedom 
is lost. It was shown in Ref.\ \cite{Litim:2014uca} that in this latter case the theory exhibits an 
interacting UV fixed point in all four couplings. This fixed point is controllable in perturbation 
theory, provided $0<\eps\ll 1$.
The existence of such an interacting UV fixed point ensures that the theory is a fundamental 
one, i.e., it is valid at arbitrarily short and large distances. Furthermore, the scalar interactions 
are free from the triviality problem because of the presence of an interacting UV fixed point.  
Therefore, the elementary scalars are part of a Wilsonian fundamental theory.  

After a lengthy study of the zeros of the theory \cite{Litim:2014uca}, using the gauge-matter 
system \eq{betag} -- \eq{betav}, and including also the investigation of the 
stability of the associated classical 
and quantum scalar potential \cite{Litim:2015iea} one arrives at the only mathematically and 
physically acceptable fixed point accessible in perturbation theory:  
\beq\label{alphaNNLO}
\begin{array}{rcl}
\alpha_g^*&=&
\frac{26}{57}\,\eps+ \frac{23 (75245 - 13068 \sqrt{23})}{370386}\,\eps^2
+{\cal O}(\eps^3)
\\[1.ex]
\alpha_y^*&=&
\frac{4}{19}\,\eps+\left(\frac{43549}{20577} - \frac{2300 \sqrt{23}}{6859}\right)\,\eps^2
+{\cal O}(\eps^3)
\\[1ex]
\alpha_h^*&=&
\frac{\sqrt{23}-1}{19}\,\eps+{\cal O}(\eps^2)
\,,\\[1.ex]
\al {v1}^*&=&
-\frac{1}{19} \left(2 \sqrt{23}-\sqrt{20 + 6 \sqrt{23}}\right)\,\eps
+{\cal O}(\eps^2)\,.
\end{array}
\eeq
Here we give the analytic expression of the fixed point in an expansion in the 
small $\epsilon$ parameter.

\begin{figure}[t]
\centering
\includegraphics[width=.8\textwidth]{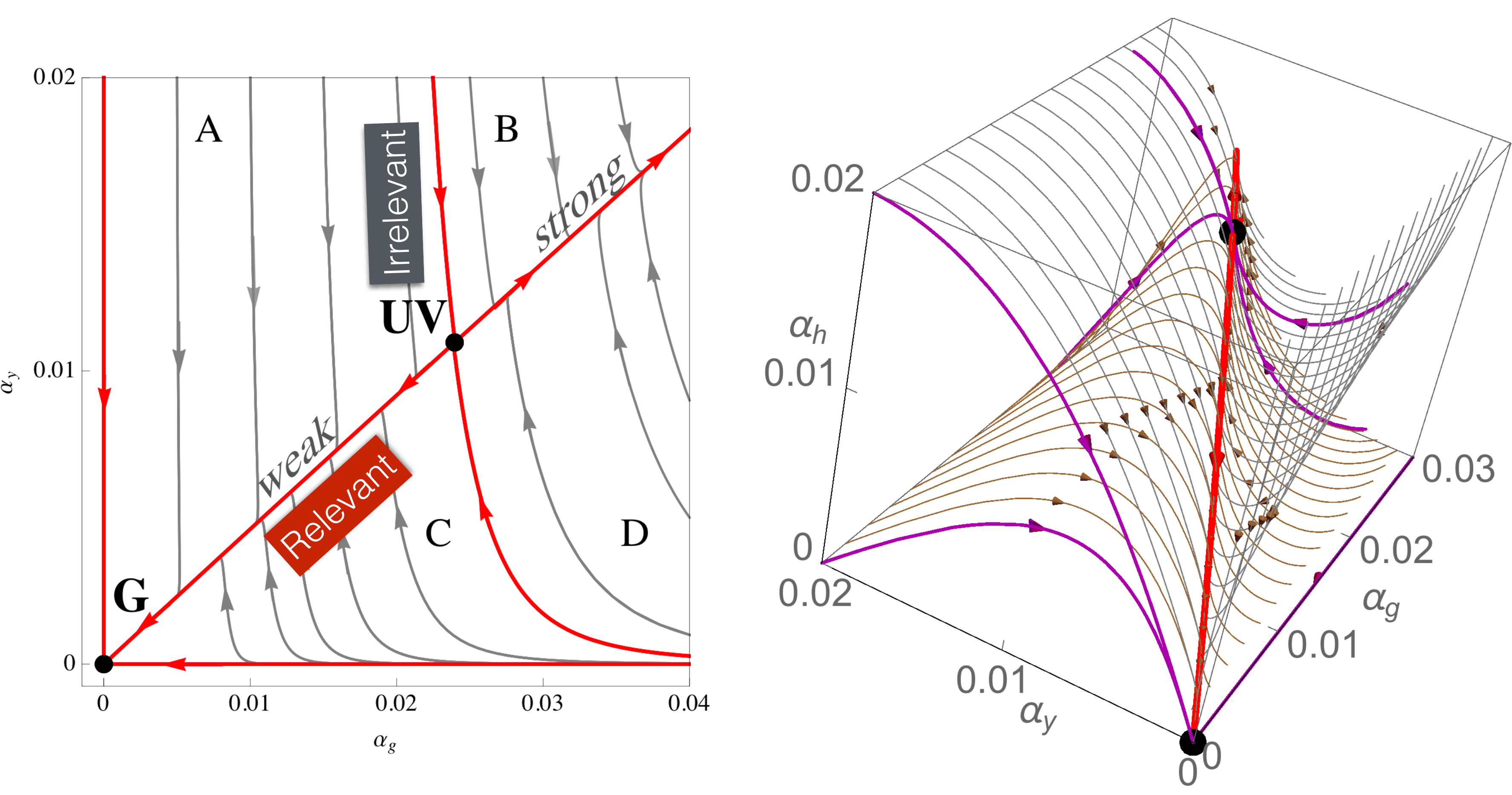}
\caption{\label{pPD} Review of the phase diagram of Ref.\ \cite{Litim:2014uca,Litim:2015iea}. 
The gauge-Yukawa subsector of couplings $(\al g,\al y)$ is shown
to leading order in the 
left-hand panel and the gauge-Yukawa-scalar subsector $(\al g,\al y,\al h)$ at 
next-to-next-to-leading order accuracy in the right-hand panel for 
$\eps=0.05$. Also shown are the UV and IR fixed points (dots), the UV-safe trajectories 
(thick red line). A few trajectories are highlighted as thin magenta lines, and a few  
generic trajectories are shown as thin gray lines. Arrows point towards the IR. 
Further details can be found in Ref.\ \cite{Litim:2014uca,Litim:2015iea}.}
\end{figure}

The phase diagram of the theory was established in Ref.\ \cite{Litim:2014uca} at 
next-to-leading order accuracy and extended to the next-to-next-to-leading
order in Ref.\ \cite{Litim:2015iea}  where the effects from the running scalar couplings 
were considered. In order to keep the paper self-contained we summarise in Fig.\ \ref{pPD} the 
phase diagram of the theory shown in Ref.\ \cite{Litim:2015iea}. In the left-hand 
panel we show the RG trajectories for the $(\al g,\al y)$ couplings,
while in the right-hand panel the three-dimensional RG flow is
illustrated, that includes also the coupling $\al h$. The two plots include the UV and IR fixed
points. We have also indicated in the left panel the relevant 
and irrelevant directions dictated by the signs of the scaling exponents. 

The IR fixed point is 
non-interacting and it is therefore located at the origin of coupling space. The thick red line 
connects the IR and UV fixed point and therefore is the UV-complete trajectory that we term 
the line of physics. The line of physics is also known as separatrix since it separates different 
regions of the theory in RG space. Along the line of physics the theory is non-interacting in the 
deep IR. The separatrix continues beyond the UV fixed point towards large couplings in the IR, 
leading to a strongly coupled theory presumably breaking conformality and chiral symmetry in 
the IR.  The region of the RG phase diagram emanating from the UV-stable fixed point 
and leading to 
stable trajectories is known as the {\it UV critical surface}. Here this critical surface is 
one-dimensional \cite{Litim:2014uca} and it has a dynamical nature.  

As mentioned above, the separatrix connects the UV fixed point with the Gaussian one and it 
agrees with the UV critical surface near the fixed point \cite{Litim:2014uca}. Although one can 
always determine numerically the globally defined separatrix, it is illuminating, in view of their 
use in the thermodynamical analysis, to consider an analytical approximation that is accurate 
in the {limit of vanishing $\epsilon$}. This leads to the following 
relations among the couplings along the separatrix \cite{Litim:2015iea}:
\beq
\label{critsurf}
\begin{array}{rcl}
\al y &=&
 \frac{6}{13}\, \al g \ , \\[2ex]
\al h & = &
\frac{3}{26} \left(\sqrt{23}-1\right)\,\al g \ , \\[2ex]
\al v & = &
\frac{3}{26} \left(\sqrt{20 + 6\sqrt{23}}-2\sqrt{23}\right)\,\al g \,.
\end{array}
\eeq
The one-dimensional nature of the line of physics is encoded in the fact that it is sufficient to 
know the running of the gauge coupling in order 
to determine the running of all the other couplings. 
The precise analytic running of the gauge coupling was determined in 
Ref.\ \cite{Litim:2015iea} and reads 
\beq\label{critg}
\alpha_g(\mu) =
\frac{\alpha_g^\ast}{1+W(\mu) }
\ ,
\eeq
where $W(\mu)\equiv W[z(\mu)]$ is the Lambert function satisfying the relation 
$z = W \exp W$ with 
\bea
\label{W}
W&=&\frac{\alpha_g^*}{\alpha_g}-1\qquad {\rm and} \qquad
z= \left(\frac{\mu_0}{\mu}\right)^{\frac{4\epsilon}{3}\alpha_g^*}
\left(\frac{\alpha_g^*}{\alpha_g^0}-1\right)
\exp\left(\frac{\alpha_g^*}{\alpha_g^0}-1\right)\,.
\eea
Here $\alpha_g^0$ is the value of the gauge coupling at the scale $\mu_0$, with $\mu/\mu_0$ 
ranging between $0$ and $\infty$ and the gauge coupling ranging between 
$0<\alpha_g^0<\alpha_g^\ast$.   

Inserting Eq.\ \eq{critg} into Eq.\ \eq{critsurf} yields an analytic description 
of the RG evolution of  
all couplings along the line of physics. This constitutes the zero-temperature information we 
need to establish the thermodynamical properties of the theory.

At asymptotically high energies $W(\mu)$ vanishes while it grows towards the infrared. It is 
convenient to fix $\alpha^0_g$ via 
$\alpha^0_g \equiv \alpha^*_g/(1+k)$ with $k \in \mathbb{R}_+$, 
which in practice amounts to fixing the arbitrary renormalization reference scale 
$\mu_0$ along the RG flow. As pointed out in Ref.\ \cite{Litim:2015iea} the value $k=1/2$, 
i.e., $\alpha^0_g = 2\alpha^*_g/3$, 
corresponds to an exact critical transition scale 
$\mu_0 = \Lambda_c$ above which the physics is dominated by the interacting 
UV fixed point and below which it is governed by the Gaussian IR fixed point.  
The interacting nature of the UV fixed point is expressed by the fact 
that it is approached as a power law in the renormalization scale 
\begin{equation}
\alpha_g (\mu ) \simeq \alpha^\ast_g + (\alpha_g^0 - \alpha_g^*) \left( \frac{\mu}{\bar{\mu}_0}
\right)^{-\tfrac{104}{171} \epsilon^2} \, ,
\end{equation}
where  $\bar{\mu}_0 =  \mu_0 (1 + \cal{O}(\epsilon))$. We have used Eq.\ (\ref{alphaNNLO})   
and that, in the deep-UV limit, the Lambert function approaches zero as
\begin{equation}
\lim_{\mu/\mu_0 \to \infty}  W(\mu) \propto 
\left( \frac{\mu}{\mu_0}\right)^{-\tfrac{104}{171}\epsilon^2}\,.
\end{equation}
 
\begin{figure}[h]
\begin{center}
\includegraphics[width=.46\textwidth]{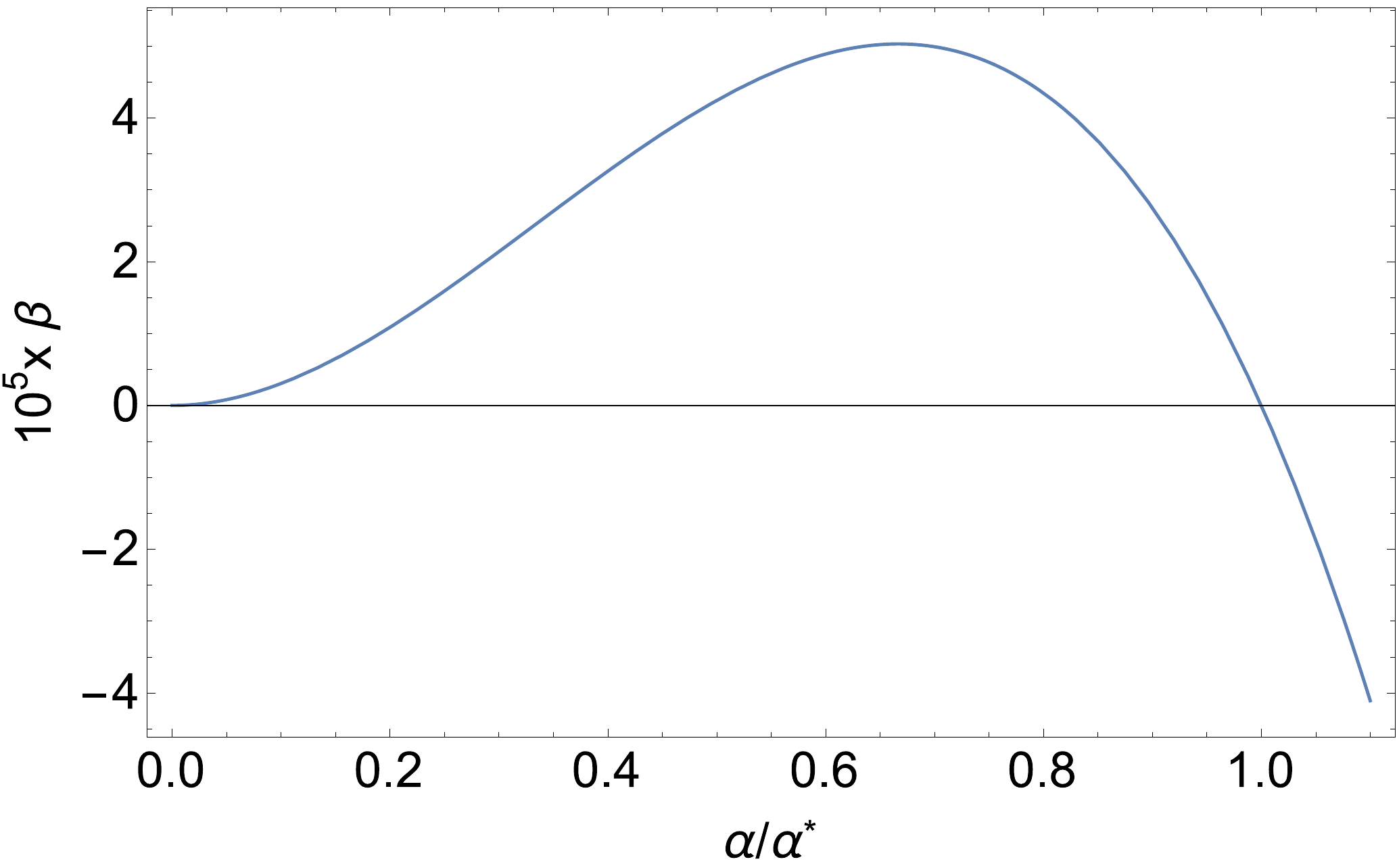} \hskip .4cm
\includegraphics[width=.46\textwidth]{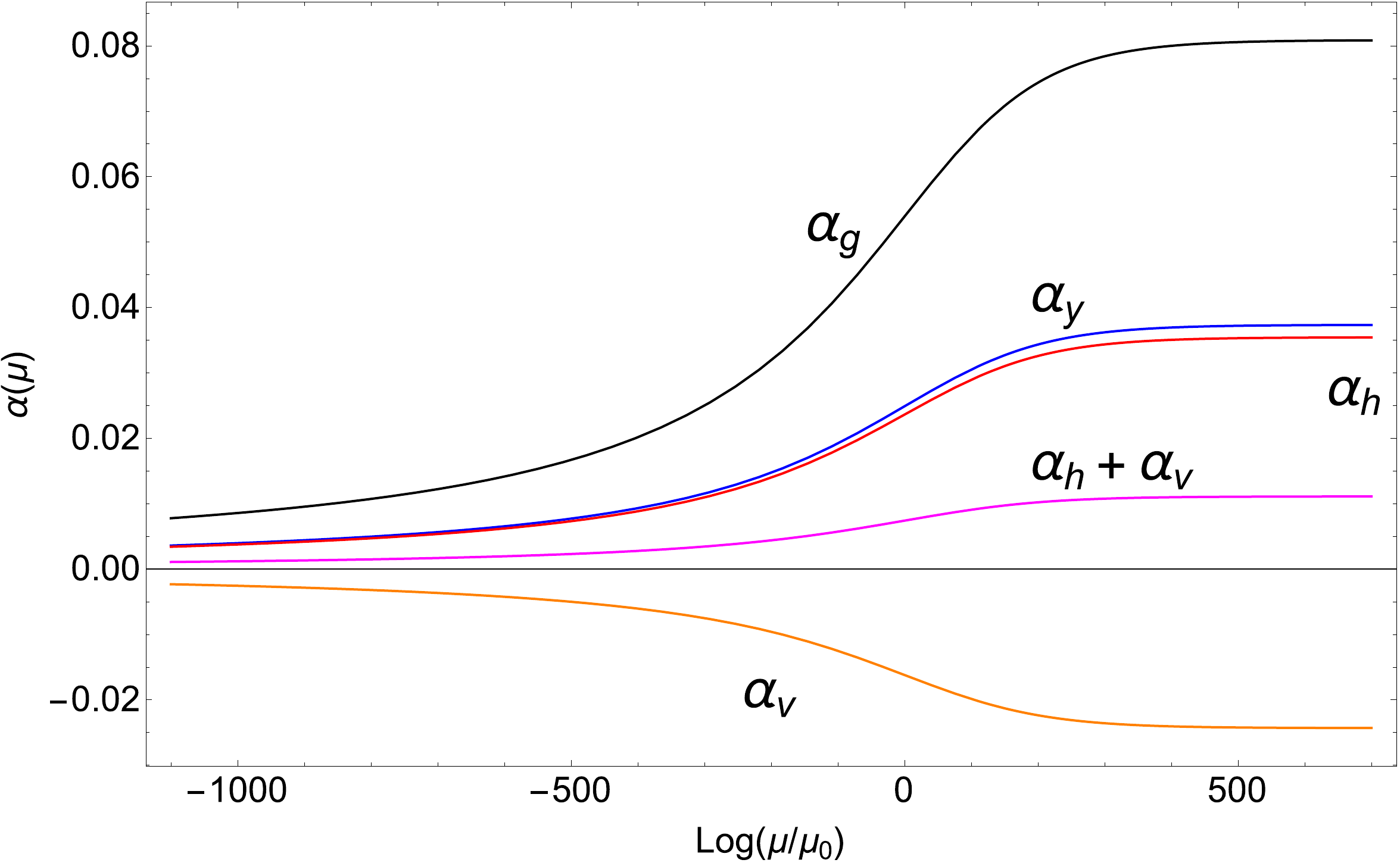} 
\caption{Left panel: Gauge beta function along the line of physics displaying the characteristic 
asymptotically safe behaviour. Right panel: From top to bottom the running of the gauge,
Yukawa, single trace, single trace plus double trace, and only single trace coupling is shown
along the line of physics.  We have chosen $k=1/2$ and $\epsilon=0.1$.}
\label{running}
\end{center}
\end{figure}

There are several nice and distinctive features of the analytically controllable and completely
asymptotically safe dynamics presented above. In particular, it constitutes
an ideal laboratory to 
investigate thermodynamical properties of the theory that for certain aspects resembles 
${\cal N}=4$ theory. One of the similarities is the fact that along the line of physics all
couplings are related. This is also a basic feature of ${\cal N} =4$ theory 
due, however, to the high degree of space-time supersymmetry. In the non-supersymmetric 
case the relations among the couplings are 
dynamical in nature being dictated by the dimension 
of the critical surface. In Fig.\ \ref{running} we show in the left panel the beta function of the 
gauge coupling along the line of physics linking the Gaussian fixed point with the interacting 
UV fixed point. In the right panel the running of all couplings along the line of physics 
is shown using Eqs.\ \eqref{critsurf} and \eqref{critg}. 
From the figure the completely asymptotically safe nature of the theory is evident. For the sake
of completeness we mention that, at fixed $N_c$ and large $N_f$, it has been 
argued \cite{Pica:2010xq} that an asymptotically safe theory can emerge without scalars 
and to leading order in $1/N_f$. Further physical properties of this intriguing possibility 
were investigated in Ref.\ \cite{Litim:2014uca}.   
 
\section{Asymptotically safe Thermodynamics}
\label{AST}
We will now study thermodynamical quantities of the theory to leading (LO), 
next-to-leading (NLO), and next-to-next-to-leading order (NNLO) in the couplings.  
The details of the computation of the thermodynamic pressure for the 
gauge-Yukawa theory, that are valid for any number of colours and flavours and 
applicable as well to the entire phase diagram of the theory, are provided in App.\ 
\ref{Appendix-T}. We  work here in the Veneziano limit and along the line of physics 
\eqref{critsurf}.

From Fig.~\ref{running} one can immediately see that there are three relevant energy regions 
with distinct dynamics: the one dominated by the Gaussian IR fixed point, i.e., 
$\mu\ll \Lambda_c$, the one dominated by the interacting UV fixed point 
$\mu\gg \Lambda_c$, and the cross-over energy region for which 
$\mu\sim\Lambda_c$. By identifying, for example, the renormalization scale 
with the temperature, at zero chemical potential, we can test the thermodynamic 
properties of the asymptotically safe plasma along the entire line of physics.  

\subsection{Hot Asymptotically Safe Pressure to Leading Order }

Near the Gaussian IR fixed point the theory is non-interacting, and the 
ideal-gas limit applies. This constitutes the LO contribution along the entire line of physics. 
Specialising the results of App.\ \ref{Appendix-T} for the pressure to 
the Veneziano limit $(N_c, N_f \gg 1)$ at zero chemical potential, 
and normalising it to the one of the gluons (in the same limit) we have: 
\begin{equation}
\label{LO}
\frac{p_0}{p_{0,g}}  = 1 + \frac{N_f^2}{N_c^2} + \frac{7}{4}\frac{N_f}{ N_c} \ .
\end{equation}
Here we notice that the ratio depends at most {quadratically} on $N_f/N_c$. 
We are, however, considering an expansion in $\epsilon = N_f/N_c - 11/2$ 
and therefore re-express the result in terms of $\epsilon$: 
\begin{equation}
\label{LO2}
\frac{p_0}{p_{0,g}}  =  \frac{327}{8}  + \frac{51}{4}  \epsilon + \epsilon^2  \ .
\end{equation}
We note that already to this order the thermodynamical expression depends at most on 
$\epsilon^2$ and that furthermore the $\epsilon$ expansion allows for a new handle on  
the thermodynamical expansion, which is absent for a generic gauge-Yukawa theory.

\subsection{Hot Asymptotically Safe Pressure to Next-To-Leading Order }

The previous LO expression for the pressure is exact at the IR fixed point because of 
its non-interacting nature. Since we choose to identify 
the renormalization scale with the
temperature, this then occurs for very small temperatures, i.e., 
$\mu \equiv T \ll \Lambda_c$. However, when the temperature rises, the plasma 
starts feeling the various interactions. To NLO the 
pressure as function of the temperature reads:
\begin{equation}
  \frac{p_{0+ 2}}{p_{0,g}} = \frac{327}{8}   +\frac{51}{4} \epsilon + \epsilon^2   - 5 
  \left[ \alpha_g  + (\alpha_v  + 2\alpha_h) \left( \frac{11}{2} + \epsilon \right)^2 \right]
 - \frac{25}{4}\left[ \alpha_g \left( \frac{11}{2} + \epsilon \right)+ \alpha_y \left( 
 \frac{11}{2} + \epsilon \right)^2\right]   \ .
\label{LO+NLO}
\end{equation}
Besides the trivial $T^4$ scaling which cancels between numerator and
denominator, there is an additional temperature dependence due to the running of the 
couplings which are evaluated at the temperature $T$. In a conformal field theory, 
however, we can only consider ratios of scales or, equivalently, the couplings 
are measured in units of a reference value. Along the line of physics the 
natural choice for the reference scale is $\Lambda_c$, corresponding to
a value of the gauge coupling, which is $2/3$ of its fixed-point value. 
This is the scale above which the physics is dominated by the UV fixed point, while below 
it is governed by the Gaussian IR one. This allows us to immediately determine the 
two limiting values of the pressure obtained for $T \ll \Lambda_c$ and for $T\gg \Lambda_c$.
 
For $T \ll \Lambda_c$ the physics is dominated, as already mentioned earlier, by the 
non-interacting fixed point and therefore the NLO pressure coincides with its
ideal-gas expression,
\begin{equation}
  \frac{p_{0+ 2}}{p_{0,g}} = \frac{327}{8}   +\frac{51}{4} \epsilon + \epsilon^2  
  = 40.875 +12.75 \, \epsilon + \epsilon^2   \ , \qquad T\ll \Lambda_c \, .
 \end{equation}
However, for  $T \gg \Lambda_c$, i.e., near or at the UV fixed point we can use the 
fixed-point values for the couplings \eqref{alphaNNLO}, which yields: 
\begin{equation}
 \frac{p_{0+ 2}}{p_{0,g}} 
= 40.875 -84.6877\, \epsilon + {\cal O}(\epsilon ^2) \ , 
\quad T\gg \Lambda_c \, .
\end{equation}
At zero temperature and chemical potential we have solved the theory to the maximum 
known order in perturbation theory that abides the Weyl consistency conditions. 
This implies that we know the gauge and Yukawa couplings to the second order in 
$\epsilon$ and the scalar couplings to the leading order in $\epsilon$. This limits 
the {expansion of the pressure in powers of $\epsilon$} to the 
{next-to-leading order}
in $\epsilon$. Interestingly, we observe a net drop of the pressure when normalized to the 
ideal-gas limit, valid in the deep IR, due to the interacting nature of the 
asymptotically safe plasma.

We can also determine the pressure along the entire line of 
physics by using Eqs.\ \eqref{critsurf} and \eqref{critg}, where Eq.\ \eqref{LO+NLO} 
reduces to
\begin{equation}
  \frac{p_{0+ 2}}{p_{0,g}} =40.875  + 12.75 \epsilon + \epsilon^2 -(213.613 
  + 69.6094 \epsilon  + 5.75995 \epsilon^2)\alpha_g(T)   \ , \qquad  {\rm for ~ any ~} T \ . 
\label{p02}
\end{equation}
We use Eq.\ \eqref{critg} and replace the renormalisation scale $\mu$ with $T$ and the 
reference scale $\mu_0$ with $\Lambda_c$ and write: 
\beq\label{critgT}
\alpha_g(T) =
\frac{\alpha_g^\ast}{1+W(T) }\ .
\eeq
From the knowledge of the functional dependence on $T$ we deduce that the  $T^4$  
coefficient of the pressure decreases monotonically when the temperature increases.  
The theory assumes ideal-gas behaviour only in the deep IR.  This is different from the 
case of asymptotically free field theories where the ideal-gas limit 
is approached in the deep UV. 
\begin{figure}[h]
\begin{center}
\includegraphics[width=.55\textwidth]{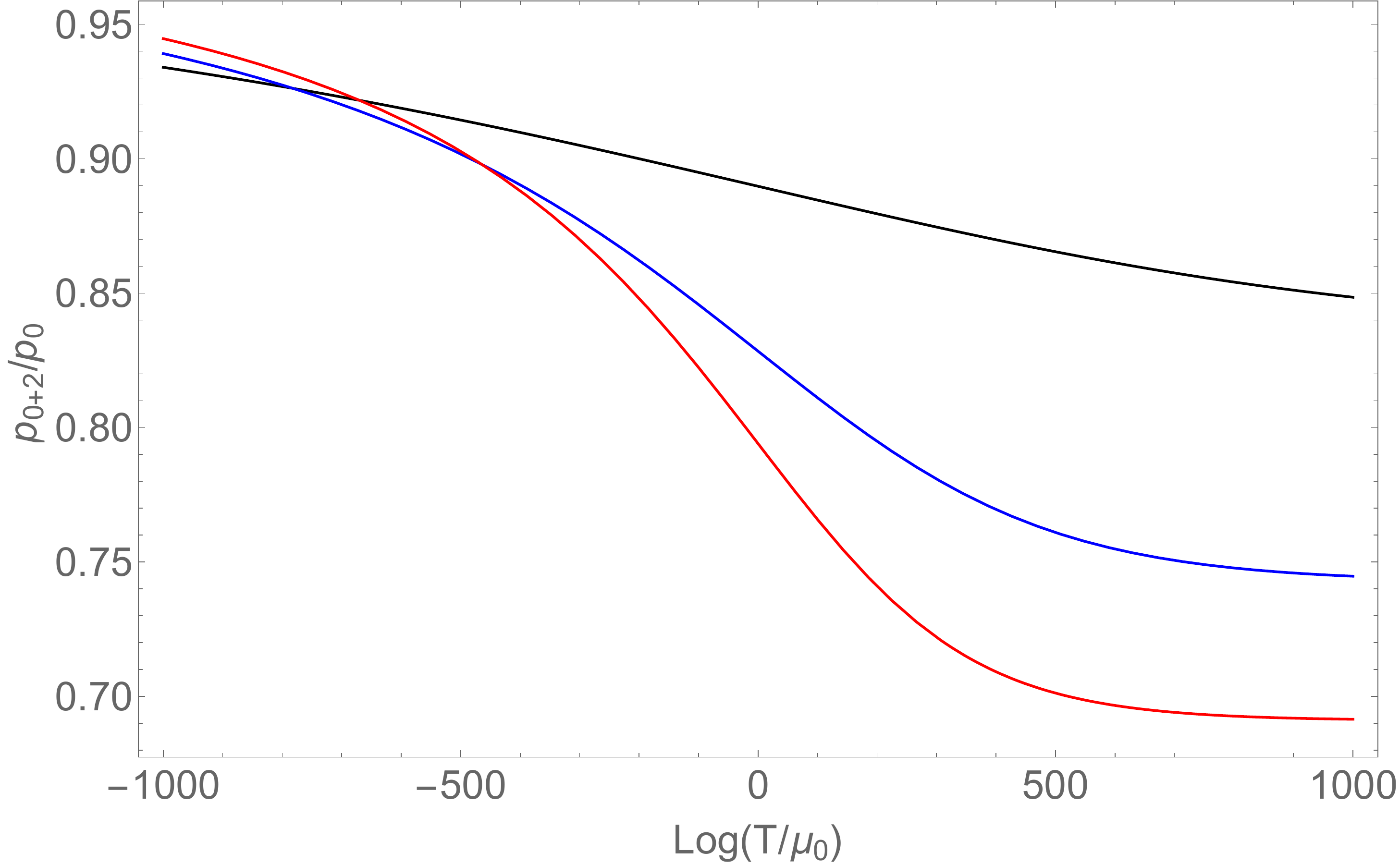} \hskip .4cm
   \caption{ Pressure normalised to the leading-order ideal-gas value, 
   up to NLO corrections, as function of the temperature.  We have chosen $k=1/2$, meaning 
   that $\mu_0 = \Lambda_c$. From bottom to top $\epsilon$ assumes the values $0.08$ (red), 
   $0.07$ (blue), and $0.05$ (black).}
\label{pressureT}
\end{center}
\end{figure}
From Fig.\ \ref{pressureT} we observe {a decrease} in the value of 
the pressure around $\Lambda_c$, now normalised to the ideal-gas limit, 
when increasing the 
temperature. By increasing $\epsilon$ the change in the pressure is more pronounced.

 \subsection{Hot Asymptotically Safe Pressure to Next-To-Next-To-Leading Order}
\begin{figure}[h]
\begin{center}
\includegraphics[width=1.\textwidth]{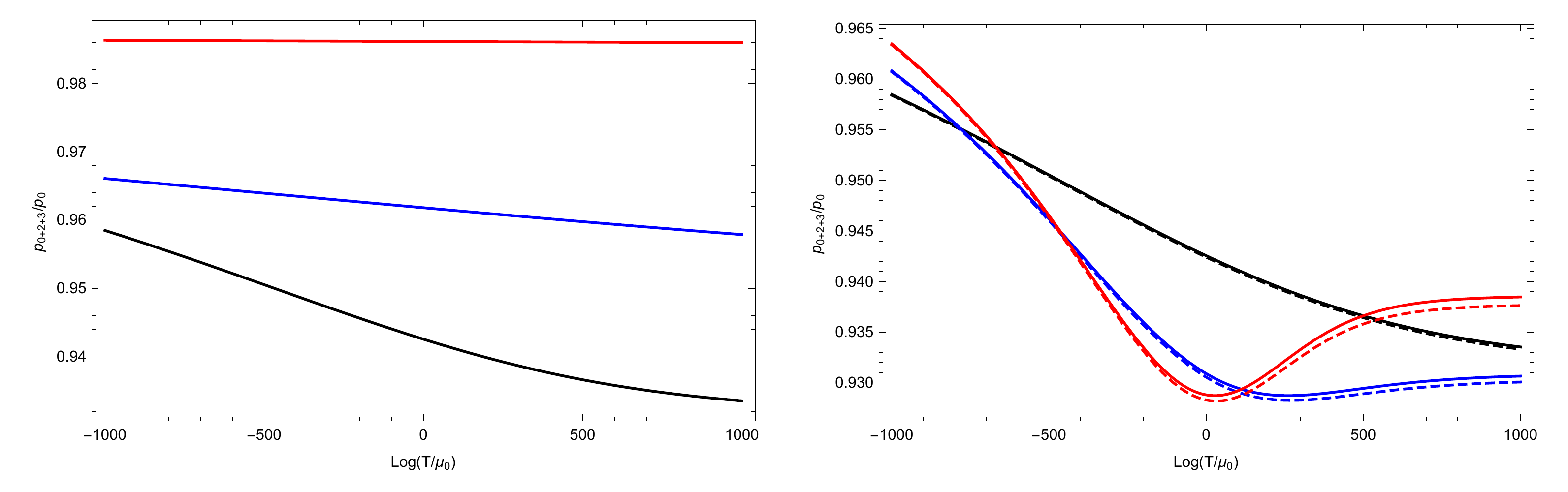}
\caption{Pressure {up to NNLO}, normalised to {its LO value}, 
as function of the temperature.  We have chosen $k=1/2$ meaning that 
$\mu_0 = \Lambda_c$.  Left panel: The values for $\epsilon$ are, from bottom to top, $0.05$ 
(black), $0.03$ (blue), and $0.01$ (red). Right panel: The {$\epsilon$} 
values for the solid curves are $0.08$ (red), $0.07$ (blue), and $0.05$ (black). We kept some 
of the higher-order corrections in $\epsilon$ for the respective dashed curves. }
\label{pressureTNNLO}
\end{center}
\end{figure}

At this order we observe the emergence of non-analytic contributions in the couplings.
More specifically, the leading contributions will start at $ {\cal O}(g^3)$ and
${\cal O} (u^{3/2}, v^{3/2})$. These come from the plasmon-ring diagrams 
\cite{Kapusta:2006pm}, cf.\ Fig.\ \ref{fig2}, and the detailed computation of their 
contribution to the pressure is given in App.\ \ref{ANNLO}. The respective contribution to the 
pressure in the Veneziano limit reads: 
\begin{equation}
\label{p3}
  \frac{p_3}{p_{0,g}} = \frac{10}{\sqrt{3 }} \alpha_g^{\frac{3}{2}} (15+2\epsilon)^{\frac{3}{2}} 
  +  \frac{5}{\sqrt{3}}   \left( 2\alpha_h + \alpha_v + \alpha_y\right)^{\frac{3}{2}}
  (11+2\epsilon)^{2} =  150\sqrt{5} \alpha_g^{\frac{3}{2}} +  \frac{605}{\sqrt{3}}   
  \left( 2\alpha_h + \alpha_v + \alpha_y\right)^{\frac{3}{2}} + {\cal O}(\epsilon^{\frac{5}{2}})\ . 
\end{equation}
Here we used the fact that all 
couplings are already of order $\epsilon$, see Eq.\ \eqref{alphaNNLO}. 
Along the line of physics 
\eqref{critsurf} this reduces to: 
\begin{equation}
  \frac{p_3}{p_{0,g}} =  704.061 \, \alpha_g(T)^{\frac{3}{2}}  + {\cal O}(\epsilon^{\frac{5}{2}})\, . 
\end{equation}
Besides the fact that the contribution is non analytic in $\epsilon$ we learn that it starts at  
${\cal O}(\epsilon^{3/2})$ and that it is positive, differently from the NLO contribution 
which is negative and starts at {$O(\epsilon)$}. 
Near the UV fixed point, i.e., at temperatures $T \gg \Lambda_c$, it assumes the value: 
\begin{equation}
    \frac{p_3}{p_{0,g}} =  216.899 \, \epsilon^{\frac{3}{2}}  + {\cal O}(\epsilon^{\frac{5}{2}})\, , 
    \qquad  {T\gg \Lambda_c} \ .
\end{equation}
The full pressure to ${\cal O}(\epsilon^{3/2})$ at nonzero temperature, zero chemical 
potential, in the Veneziano limit, and along the line of physics reads: 
\begin{equation}
  \label{NNLO-a}
  \frac{p_{0+ 2+3}}{p_{0,g}} =40.875  + 12.75 \epsilon   -213.613\, \alpha_g(T)  + 704.061 \, 
  \alpha_g(T)^{\frac{3}{2}} + {\cal O}(\epsilon^{2})\ , \qquad  {\rm for ~ any ~} T  \ . 
\end{equation}
The pressure normalised to the non-interacting limit is shown in the two panels of Fig.\
\ref{pressureTNNLO} for different values of $\epsilon$. 
For the plot on the left the values for $\epsilon$ are, from bottom to top, $0.05$ (black), 
$0.03$ (blue), and $0.01$ (red), while the values for the solid curves on the right are 
$0.08$ (red), $0.07$ (blue), and again, for reference, $0.05$ (black). For the dashed curves
we retain some of the higher-order corrections in $\epsilon$ in order
to gain an estimate on their order of magnitude. These come by retaining all 
the {powers} in $\epsilon$ from Eqs.\ \eqref{p02} and \eqref{p3}. 
Overall the analysis shows that we have 
good control of the perturbative expansion up to NNLO terms. 
Because the {NNLO} corrections to the pressure are positive, the 
{total value of the} pressure increases with respect to {its NLO
value}, but not {with respect to its LO value}, when approaching the 
UV fixed point. Furthermore, the normalised pressure starts developing a minimum near 
$\Lambda_c$ when $\epsilon$ increases above the value $0.05$.  

The decrease of the pressure normalised to the one of the {ideal} gas is not 
guaranteed to be monotonous because of the effects of the NNLO corrections. However,
within the realm of perturbation theory this quantity globally decreases, i.e.,
\begin{equation}
\Delta p_{\rm norm} \equiv \frac{p_{0+ 2+3}(T\ll \Lambda_c)}{p_{0}} 
-  \frac{p_{0+ 2+3}(T\gg \Lambda_c)}{p_{0}} = 2.384 \, \epsilon 
- 5.306 \, \epsilon^{\frac{3}{2}} \geq 0\ ,
\end{equation}
provided that $\epsilon < 0.202$. This is guaranteed by the fact that the radius of convergence 
of the expansion is, at zero temperature, $\epsilon<0.11$ \cite{Litim:2014uca}, and it is even smaller at nonzero 
temperature. 
 
\section{Asymptotically safe entropy}
\label{entropy}
Another important quantity to determine is the entropy density of the system, which 
is related to the pressure via 
\begin{equation}
s= \frac{dp}{dT} \ .
\end{equation}
Given that in the present system, and along the line of physics, the pressure can be written 
as a function of only one coupling we have: 
\begin{equation}
p = f\left ( \alpha_g(T)\right) \frac{\pi^2}{90}T^4 \ . 
\end{equation}
In the non-interacting gas $f(\alpha_g(T))$ is the number of {boson
degrees of freedom plus $7/4$} times the number of Weyl fermions. The entropy 
{density} normalised to the one of {an ideal} gas of gluons reads: 
\begin{equation}
\frac{s}{s_{0,g}} =\frac{1}{2(N^2_c  - 1)} \left[ f +  \frac{\beta(\alpha_g)}{4}  
\frac{d f}{d \alpha_g} \right] =  \frac{p}{p_{0,g}} 
+  \frac{\beta(\alpha_g)}{4}  \frac{d (p/p_{0,g})}{d \alpha_g}   
=   \frac{p}{p_{0,g}} +  \frac{1}{4} \frac{d (p/p_{0,g})}{d \ln T} \  ,
\label{normalisedS}
\end{equation}
with $\beta(\alpha_g) = { d \alpha_g}/{ d \ln \mu}$, where $\mu = T$, 
is the gauge beta function along the line of physics. We have used the fact that 
$f = 2(N^2_c  -1 ) p/p_{0,g}$. Because the beta functions vanish at a fixed point we 
have that at the IR and UV fixed points the normalised entropy {density} 
agrees with the normalised fixed-point pressure and therefore: 
\begin{eqnarray}
\frac{s_{IR}}{s_{0,g}}   &= &   \lim_{T/\Lambda_c \rightarrow 0 } \frac{p}{p_{0,g}} 
=  \frac{f_{IR}}{2N_c^2 }= 1 + \frac{N_f^2}{N_c^2}  + \frac{7}{4}\frac{N_f}{N_c} 
= 40.875 + 12.75 \epsilon + \epsilon^2 \ , \\ 
 \frac{s_{UV}}{s_{0,g}} & = &  \lim_{T/\Lambda_c \rightarrow \infty}   \frac{p}{p_{0,g}} 
 = \frac{f_{UV}}{2N_c^2} =    \frac{f({\alpha_g^{\ast})}}{2N_c^2 } 
 = 40.875 - 84.6877 \epsilon +216.899 \epsilon^{3/2} + {\cal O}(\epsilon^2)  \ . 
\end{eqnarray}
Away from the fixed points the normalised entropy {density} 
and pressure differ by the quantity 
\begin{equation} 
\frac{s}{s_{0,g}}  - \frac{p}{p_{0,g}} =   \frac{\beta(\alpha_g)}{4}  
\frac{d (p/p_{0,g})}{d \alpha_g} \ ,
\end{equation}
which is directly proportional to the beta function of the theory. This behaviour is different, 
for example, from the {case of ${\cal N} =4$ theory where} the beta 
function vanishes identically. 
  
Using the NNLO expression for the pressure from Eq.\ \eqref{NNLO-a} we deduce
\begin{equation} 
\frac{s_{0+2+3}}{s_{0,g}}  - \frac{p_{0+2+3}}{p_{0,g}} =   -53.4033\, 
\frac{d \alpha_g(T)}{ d  \ln T} \left[ 1 - 4.94395 \, \alpha_g(T)^{\frac{1}{2}} \right] \ .
\end{equation}
This contribution being directly proportional to the gauge beta function along the line of physics 
from Fig.~\ref{running} it is clear that {it} is suppressed compared to the term 
directly proportional to the normalised pressure and therefore decreases from the IR to the UV.  
\begin{figure}[h]
\begin{center}
\includegraphics[width=.55\textwidth]{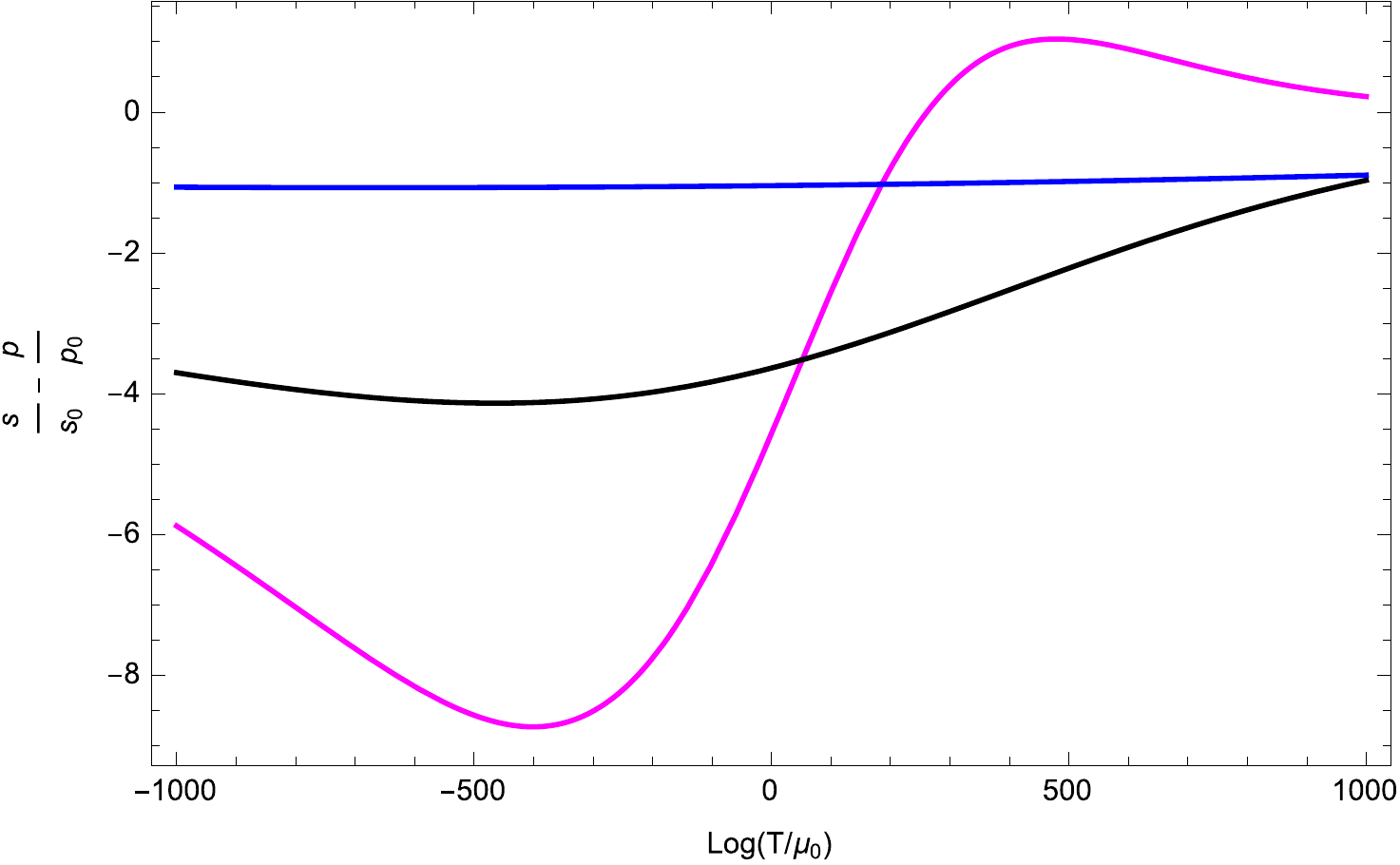} 
\caption{ We show the difference between the entropy {density} 
normalised to {its leading-order ideal-gas value} 
minus the similarly normalised pressure, 
in units of $10^{-6}$, up to NNLO corrections as function of the temperature.  
We have chosen $k=1/2$ meaning that $\mu_0 = \Lambda_c$ and from bottom to top 
$\epsilon$ assumes the values $0.07$ (magenta), $0.05$ (black), and $0.03$ (blue).}
\label{entropy-pressure}
\end{center}
\end{figure}
In Fig.~\ref{entropy-pressure} we show the difference in entropy density and 
pressure both normalised with respect to the ideal-gas limit, rather than 
the {ideal} gas of gluons. The difference is simply an overall numerical factor. 

We learn that the entropy density 
normalised to the ideal-gas limit decreases overall 
from the IR to the UV  with the dominant piece, in perturbation theory, given by the 
normalised pressure. However it does not decrease monotonically.

 \section{Asymptotically safe thermal degrees of freedom}
 \label{TDoF}
 The free energy {density} ${\cal F}(T) = - p(T)$ can be used to count 
 the physical 
 degrees of freedom of the theory at different energy scales. The temperature probes 
 the relevant degrees of freedom by  exciting them. Therefore the function 
 \begin{equation}
 f(T) = - \frac{{\cal F}(T)}{T^4} \frac{90}{\pi^2} = \frac{p(T)}{T^4} \frac{90}{\pi^2}  \ ,
 \end{equation}
 is a possible candidate to count these degrees of freedom. Alternatively one can use the 
 $T^3$ coefficient of the opportunely normalised entropy density, i.e., 
 $\displaystyle{f +{df}/{4\,d\ln T}}$. The two definitions coincide at fixed points. 
 We can now count, for the first time, the thermal degrees of freedom along the entire 
 line of physics of a completely asymptotically safe field theory. Up to an overall 
 normalisation the result is the one presented in Fig.\ \ref{pressureTNNLO}. It shows that 
 the thermal degrees of freedom decrease from the infrared to the ultraviolet, albeit not 
 monotonically.  In the deep infrared, i.e., for the cold field theory, we have of course the 
 ideal gas result 
 \begin{equation}
 f_{IR} =  \lim_{T \rightarrow 0} f(T) =  2(N^2_c - 1) + 2 N_f^2 + \frac{7}{2}  N_f N_c  \ , 
  \end{equation}
 that in the Veneziano limit reads
 \begin{equation}
 \frac{f_{IR}}{2N_c^2} = 1 + \frac{N_f^2}{N_c^2}  + \frac{7}{4} \frac{N_f}{N_c} \ . 
 \end{equation}
 By construction this function coincides with Eq.\ \eqref{LO} and provides the 
 overall normalisation. 
 
 Interestingly we discover  
 \begin{equation}
 f_{IR}  \geq f_{UV} \ , \end{equation} 
 for the classes of asymptotically safe theories investigated here with 
 \begin{equation}f_{UV} =  \lim_{T \rightarrow \infty} f(T)\ .\end{equation} 
 This demonstrates that  the inequality $f_{IR}  \leq f_{UV}$ that has been conjectured  
 to be valid for asymptotically free field theories \cite{Appelquist:1999hr} does not apply 
 to asymptotically safe field theories\footnote{The inequality has not been proven even 
 for asymptotically free theories, but it was shown to be consistent with known results and 
 then used to derive constraints for several strongly coupled, vector-like gauge theories. 
 The correct counting of the infrared degrees of freedom, with respect to the inequality, 
 for the important case of an $SU(2)=Sp(2)$ gauge theory with fermions in the 
 fundamental representation  was first performed correctly in Ref.\ \cite{Sannino:2009aw}. 
 The conjecture has been used also for chiral gauge theories 
 \cite{Appelquist:1999vs,Appelquist:2000qg}. }.  
 The fact that this function decreases in the present case is due to the fact that the 
 theory becomes interacting in the UV while it is free from interactions in the deep infrared.  
 On the other hand the a-theorem is satisfied as shown in Ref.\ \cite{Antipin:2013pya}. 
 
  \section{Conclusions}
\label{conclusions}

We have computed relevant thermodynamic properties of non-supersymmetric 
four-dimensional completely asymptotically safe field theories \cite{Litim:2014uca} 
up to NNLO. Because of the completely asymptotically safe nature of the theories that have
been investigated here the coupling constants freeze at a constant and calculable value in 
the UV. Furthermore, their value can be made arbitrarily small in the Veneziano limit because 
of the existence of a continuous control parameter. This has justified a perturbative 
determination of the vacuum and in-medium properties of the theories investigated here. 
In this work we have determined the pressure and entropy 
density of these theories to
next-to-next-to-leading order. We find that because of the nature of the interactions, 
both the pressure and the entropy density, normalised to their 
respective non-interacting ideal-gas values decrease when going from the 
infrared to the ultraviolet.  

After this initial investigation several novel avenues can be explored 
such as the response of completely asymptotically safe theories to the introduction of different 
kinds of chemical potentials. Here, we would just like to remark on one 
interesting physical phenomenon. For the sake of simplicity, we
assume that there is a single chemical potential $\mu_q$ associated
to fermion number conservation. Similarly to QCD, at nonzero $\mu_q$ and
sufficiently low temperature, the theory features a colour-superconducting phase, 
because of attractive one-gauge field exchange interactions near the Fermi surface 
\cite{Alford:2007xm}. 
Because of the Pauli principle, fermion Cooper pairs must form in channels
which are totally antisymmetric in colour-flavour-spin space. To be definite, 
let us focus on Cooper pairs in the antisymmetric spin-zero channel.
Such pairs must then be either completely antisymmetric or completely symmetric
in both colour and flavour. In QCD, the one-gluon exchange interaction is attractive in
the antisymmetric colour-antitriplet channel, which requires also an 
antisymmetric wavefunction
in flavour space. In principle, this is different in the gauge-Yukawa theories studied here: the
scalar fields can mediate attractive interactions also for symmetric representations
in colour space, which in turn demands a symmetric flavour wavefunction.
However, whether this actually happens requires a more quantitative study, since a repulsive
interaction in the symmetric colour channel may destroy the pairing.
Nevertheless, the phenomenon of colour superconductivity 
(in an antisymmetric colour channel) in these theories is robust: 
{\it any} attractive interaction, no matter how small, will destabilize the Fermi surface and
lead to the formation of Cooper pairs. Therefore, Cooper pairs will form except right at the
Gaussian IR fixed point, i.e., at $T=\mu=0$. In the perturbative regime, i.e., 
for $\epsilon \ll 1$, the system is a BCS superconductor, with a gap which is
exponentially small in the coupling. For $\epsilon \ll 1$, i.e., $N_c \sim 2N_f/11$, 
a chiral-density wave phase is not expected to occur; this would
require $N_c \gtrsim 1000 N_f$ \cite{Shuster:1999tn}.

\section*{Acknowledgments}
D.H.R.\ acknowledges the hospitality of the heavy-ion theory group at the University of
Jyv\"askyl\"a, the Helsinki Institute of Physics, {and the 
University of South Denmark at Odense,} where part of this work was done. 
The work of F.S.\ is partially supported by the Danish National Research Foundation 
under the grant DNRF:90. 

\appendix

\section{Beta functions}
\label{beta}
In the large-$N$ limit, the perturbative renormalisation group equations for the 
couplings \eq{couplings} have been obtained in Ref.\ \cite{Antipin:2013pya} in 
dimensional regularisation, also using the results of Refs.\ 
\cite{Machacek:1983tz,Machacek:1983fi,Machacek:1984zw}. In terms of 
Eq.\ \eq{eps} they are given by 
\begin{eqnarray}
\label{betag}
\beta_g&=& \frac{4}{3}\eps\,\alpha_g^2  + 
\left[ \left(25+\frac{26}{3}\eps\right) \alpha_g -\frac{1}{2} \left(11+2\eps\right)^2 \alpha_y
\right] \alpha_g^2 \\
&& + \left[ \left(\frac{701}{6}+  \0{53}{3} \eps - \0{112}{27} \eps^2\right) \alpha_g^2 - 
   \0{27}{8} (11 + 2 \eps)^2 \alpha_g \alpha_y 
   + \0{1}{4} (11 + 2 \eps)^2 (20 + 3 \eps) \alpha_y^2\right] \alpha_g^2\, ,
   \nonumber\\
\beta_y&=& \alpha_y\, \left[ (13 + 2 \eps) \,\alpha_y-6\,\alpha_g \right] \\ \nonumber
&&+\alpha_y \left[
\0{20 \eps-93}{6}\alpha_g^2 
+  (49 + 8 \eps) \alpha_g \alpha_y
-    \left(\0{385}{8} + \0{23}{2} \eps + \0{\eps^2}{2}\right) \alpha_y^2 
-  4(11 + 2 \eps) \alpha_y \alpha_h 
+ 4 \alpha_h^2\right] \, ,
\label{betay}
\\
\label{betah}
  \beta_h&=&
 -(11+ 2\eps) \,\alpha_y^2+4\alpha_h(\alpha_y+2\alpha_h)\,,\\
\label{betav}
  \beta_v&=&
12 \alpha_h^2  +4\al v \left(\alpha_v+ 4 \alpha_h+\alpha_y\right)\, ,
\end{eqnarray}
for $\beta_g,\beta_y,\beta_h$, and $\beta_v$ up to $(3,2,1,1)$-loop order, respectively. 
In the terminology of Ref.\ \cite{Litim:2014uca} we refer to this as the next-to-next-to-leading 
order (NNLO) approximation. The NLO approximation corresponds to the approximation 
in which the $(2,1,0,0)$-loop terms for $\beta_g,\beta_y,\beta_h$, and $\beta_v$ are retained. 
As discussed in Ref.\ \cite{Litim:2014uca}, this ordering of perturbation theory is also 
favoured by the Weyl consistency conditions 
\cite{Jack:1990eb,Antipin:2013pya,Jack:2014pua}. 

\section{  Explicit computation of the Thermodynamical Properties}
\label{Appendix-T} 
{In this appendix we} determine the thermodynamical quantities of the theory 
to leading (LO) and 
next-to-leading order (NLO) in the couplings. For a perturbative calculation of 
the pressure we need to identify the interaction vertices in Eq.\ (\ref{Lag}).
The gauge sector will be analogous to QCD, so we shall simply use the results from
Ref.\ \cite{Kapusta:2006pm}. For the Yukawa interaction and the self-interaction of
the scalar, however, we will be more explicit.
We therefore decompose the Yukawa interaction term $\sim y$ in Eq.\ (\ref{Lag}) with the
help of Eq.\ (\ref{H}) and the definition of the projectors onto right- and left-handed
chirality, $P_{R,L} \equiv (1 \pm \gamma_5)/2$,
\begin{equation}
\bar{\psi}_L \, H \, \psi_R + \bar{\psi}_R \, H^\dagger \psi_L = \bar{\psi} T_a \psi \, S_a
+ i \bar{\psi} T_a \gamma_5 \psi \, P_a\;.
\end{equation}
In order to compute the self-interaction terms $\sim u, v$ of the 
scalar field $H$, we utilize the decomposition (\ref{H}), the orthogonality relation
\begin{equation}
{\rm Tr} \left( T_a T_b \right) = \frac{1}{2}\, \delta_{ab}\;,
\end{equation}
and the (anti-)commutation relations for the generators of $U(N_f)$,
\begin{eqnarray}
\left\{ T_a, T_b \right\} & = & d_{abc}\, T_c\;, \nonumber \\
\left[ T_a, T_b \right] & = & i f_{abc}\, T_c\;,
\end{eqnarray}
where $d_{abc}$ ($f_{abc}$) are the totally (anti-)symmetric structure constants of
$U(N)$. We obtain
\begin{eqnarray}
{\rm Tr} \left( H^\dagger H \right) & = & \frac{1}{2} \left( S_a^2 + P_a^2 \right)\;, \label{Tr} \\
{\rm Tr} \left(H^\dagger H \right)^2 & = & \frac{1}{24} \left( d_{abn} d_{cdn} 
+ d_{acn} d_{bdn} + d_{adn} d_{bcn} \right) 
\left( S_a S_b S_c S_d + P_a P_b P_c P_d \right) \nonumber \\
&  + &  \;\frac{1}{4} \left( d_{abn} d_{cdn} + f_{acn} f_{bdn} + f_{adn} f_{bcn} \right) 
S_a S_b P_c P_d \;. \label{Tr2}
\end{eqnarray}

\subsection{Pressure to LO}

To LO, the pressure of the theory (\ref{Lag}) is that of an ultrarelativistic ideal 
gas of $N_c^2-1$ gauge fields, $N_c N_f$ Dirac fermions, and $2 N_f^2$ scalars. At
temperature $T$, and if we assume
a common chemical potential $\mu_q$ for all fermions (associated to net-fermion number 
conservation), we have
\begin{equation}
p_0(T,\mu_q) = p_{0,g}(T) + p_{0,f}(T,\mu_q) + p_{0,H}(T)\;,
\end{equation}
where \cite{Kapusta:2006pm}
\begin{eqnarray}
p_{0,g}(T) & = & 2 (N_c^2-1)\, \frac{\pi^2}{90}\, T^4\;, \nonumber \\
p_{0,f}(T, \mu_q) & = & 2 N_c N_f \left( \frac{7}{4}\, \frac{\pi^2}{90}\, T^4 + \frac{\mu_q^2 T^2}{12}
+ \frac{\mu_q^4}{24 \pi^2} \right)\;, \nonumber\\
p_{0,H}(T) & = & 2 N_f^2\,\frac{\pi^2}{90}\, T^4\;.
\end{eqnarray}

\subsection{Pressure to NLO}

\begin{figure}[h]
\begin{center}
\includegraphics[width=15cm]{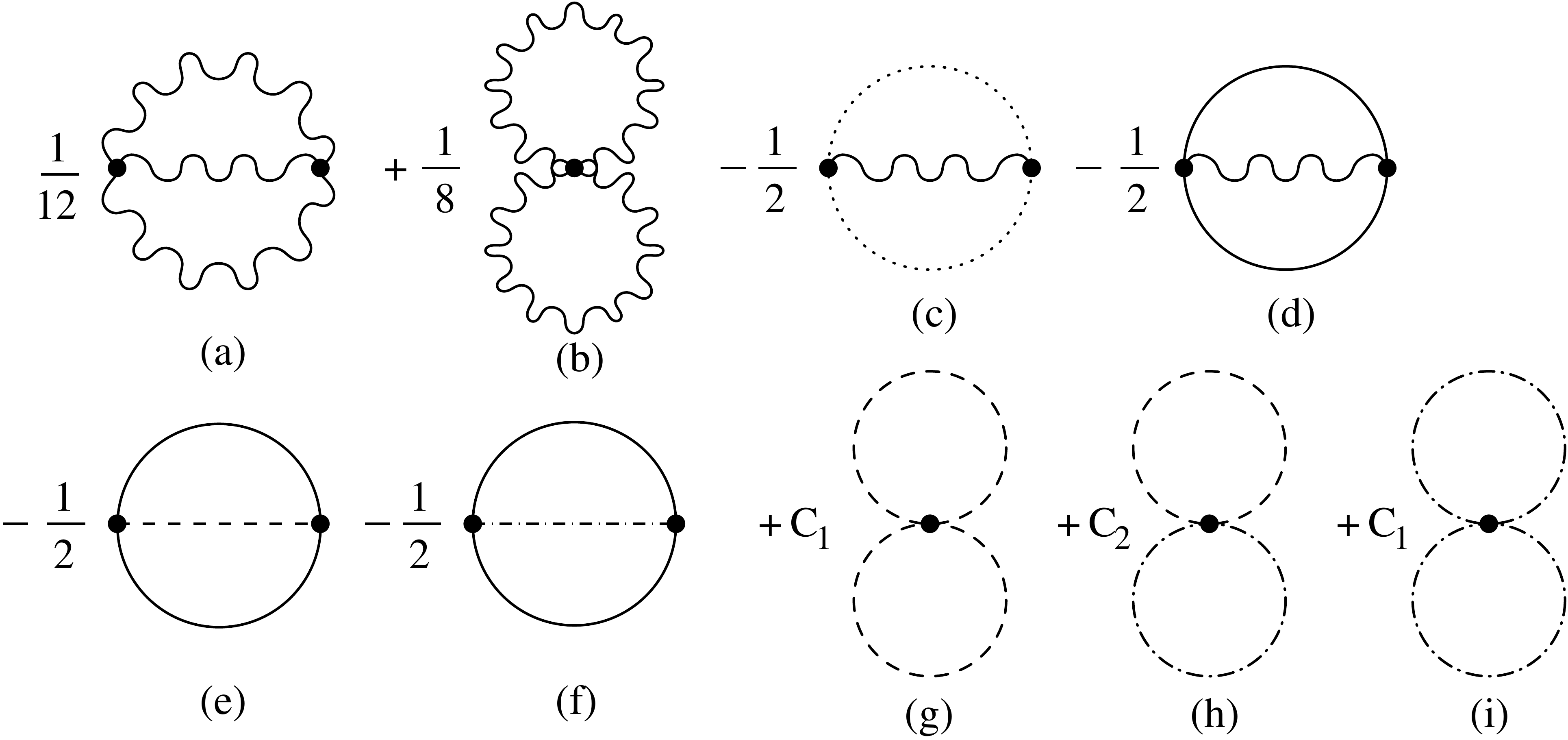}
\caption{Two-loop contributions to the pressure. Diagrams (a--d) are
the same as in QCD; gauge fields are represented by wavy lines, ghosts by dotted lines, and
fermions by solid lines. Diagrams (e,f) are the two-loop contributions in a 
Yukawa theory where fermions interact with scalar (dashed line) 
and pseudoscalar
fields (dash-dotted line). Diagrams (g--i) are the two-loop contributions arising from
the self-interaction of the scalar field $H$, decomposed in terms of
scalars and pseudoscalars. The combinatorial factors $C_{1}$ and $C_{2}$
are implicitly computed in the text, but will not be needed explicitly.}
\label{fig1}
\end{center}
\end{figure}

The NLO contribution to the pressure,
\begin{equation} \label{p2}
p_2(T,\mu_q) = p_{2,g}(T) + p_{2,gf}(T,\mu_q) + p_{2,Hf}(T,\mu_q) + p_{2,H}(T)\;,
\end{equation}
has a diagrammatic representation in terms of the two-loop diagrams shown in Fig.\ \ref{fig1}.
The first term in Eq.\ (\ref{p2}) is the contribution from the self-interaction
of the gauge fields and gauge fields with ghosts, cf.\ Figs.\ \ref{fig1}(a--c). It reads
\cite{Kapusta:2006pm}
\begin{equation} \label{p2g}
p_{2,g}(T) = -  g^2 N_c \,(N_c^2-1)\, \frac{T^4}{144} \; .
\end{equation}
The second term in Eq.\ (\ref{p2}) is the contribution from the fermion loop, Fig.\ \ref{fig1}(d),
where the fermion interacts with a gauge field. This reads \cite{Kapusta:2006pm}
\begin{equation} \label{p2gf}
p_{2,gf}(T,\mu_q) = - g^2 \,(N_c^2-1)N_f\, \left( \frac{5T^4}{576} + \frac{\mu_q^2 T^2}{32 \pi^2}
+ \frac{\mu_q^4}{64 \pi^4} \right)\;.
\end{equation}
The third term in Eq.\ (\ref{p2}) is the contribution from the fermion loop, Figs.\ \ref{fig1}(e,f),
where the fermion interacts either with a scalar or a pseudoscalar field. It turns
out that both contributions are identical. The calculation proceeds analogous to the
one for the diagram \ref{fig1}(d). The result is, up to a prefactor, identical to Eq.\ (\ref{p2gf}),
\begin{equation} \label{p2Hf}
p_{2,Hf}(T,\mu_q) = - y^2 \, N_f^2 N_c\,  \left( \frac{5T^4}{576} + \frac{\mu_q^2 T^2}{32 \pi^2}
+ \frac{\mu_q^4}{64 \pi^4} \right)\;.
\end{equation}
The last term in Eq.\ (\ref{p2}) receives contributions from the vertices $\sim u$ and
$\sim v$ in Eq.\ (\ref{Lag}),
\begin{equation} \label{p2H}
p_{2,H}(T) = p_{2,u}(T) + p_{2,v}(T)\;.
\end{equation} 
We first compute the latter. With the help of Eq.\ (\ref{Tr}) we write
\begin{equation} \label{vTr}
- v \, {\left[ {\rm Tr} \left(H^\dagger H\right) \right]^2} =
- \frac{v}{4} \left( S_aS_a S_b S_b + 2 S_a S_a P_b P_b + P_a P_a P_b P_b \right)\;.
\end{equation}
In order to produce a double-bubble diagram of the type shown in Fig.\ \ref{fig1}(g),
either (1) we can tie a leg $\sim S_a$ together with the other leg $\sim S_a$ (then we
must tie $S_b$ together with $S_b$), or (2) we can tie a leg $\sim S_a$ together with one of
the two legs $\sim S_b$  (then the other leg $\sim S_a$ must be tied together with the other
leg $\sim S_b$). Case (1) corresponds to a Hartree-type routing of internal indices and
produces an overall factor of $N_f^4$, because there are $N_f^2$ scalar fields running 
in each of the two loops. Case (2) corresponds to a Fock-type routing of internal
indices and produces an overall factor of $N_f^2$, because all indices are tied together
in a way that there is effectively only one loop in the index $a$.
Overall, we obtain a factor of $N_f^2 (N_f^2+2)$.
The same can be repeated for the pseudoscalar contribution, Fig.\ \ref{fig1}(i), with the same
result, as nothing distinguishes the two types of fields in the absence of chiral symmetry
breaking. The remaining diagram is the
one with one scalar and one pseudoscalar loop, cf.\ Fig.\ \ref{fig1}(h). Here, the 
factor is (with the factor 2 from Eq.\ (\ref{vTr})) simply $2 N_f^4$. Each double-bubble diagram 
is proportional to the square of
a tadpole which, for massless particles, has the value $T^2/12$ \cite{Kapusta:2006pm}.
Altogether we obtain
\begin{equation} \label{p2v}
p_{2,v}(T) = -v N_f^2 (N_f^2+1) \frac{T^4}{144}\;.
\end{equation}
We now compute $p_{2,u}(T)$. We first consider the contribution to the diagrams 
in Figs.\ \ref{fig1}(g) and (i), corresponding to the first line in Eq.\ (\ref{Tr2}). The three different
ways to tie legs together to form a double-bubble diagram
can be written in terms of a combination of Kronecker-deltas,
so that the prefactor becomes
\begin{equation} \label{20}
\frac{1}{24} \left( d_{abn} d_{cdn} + d_{acn} d_{bdn} + d_{adn} d_{bcn} \right)
\left(\delta_{ab} \delta_{cd} + \delta_{ac} \delta_{bd} + \delta_{ad} \delta_{bc} \right)
= \frac{1}{8} \left(d_{aan} d_{bbn} + 2 d_{abn} d_{abn} \right)
= \frac{1}{4} \, N_f (2N_f^2+1)\;,
\end{equation}
where we have used $d_{aan} = \sqrt{2} N_f^{3/2} \delta_{n0}$ (a sum over $a$ is implied) 
and $d_{abn} d_{abn} = N_f (N_f^2 +1)$. (This can be proven using the
relationship $d_{ijk} d_{ijk} = (N_f^2-1) (N_f^2-4)/N_f$, $i,j,k=1, \ldots, N_f^2-1$, 
for the symmetric structure constants of $SU(N_f)$.) Now we consider the contribution to
the diagram in Fig.\ \ref{fig1}(h), arising from the second line in Eq.\ (\ref{Tr2}). 
Here, there is only one way to tie the legs together,
\begin{equation} \label{21}
\frac{1}{4} \left( d_{abn} d_{cdn} + f_{acn} f_{bdn} + f_{adn} f_{bcn} \right) \delta_{ab}
\delta_{cd} = \frac{1}{4} \left( d_{aan} d_{ccn} + 2 f_{acn} f_{acn} \right)
= \frac{1}{2} \, N_f (2 N_f^2-1)\;,
\end{equation}
where we have used the relationship $f_{abn} f_{abn} \equiv f_{ijk} f_{ijk} = N_f (N_f^2-1)$.
Putting all this together (remembering that Eq.\ (\ref{20}) is multiplied by a factor of 2, for
scalar and pseudoscalar contributions), we obtain
\begin{equation} \label{p2u}
p_{2,u}(T) = - \frac{u}{2} \, N_f (2N_f^2+1 + 2N_f^2-1) \frac{T^4}{144}
= - 2u\,N_f^3\, \frac{T^4}{144}\;.
\end{equation}
Adding the contributions (\ref{p2g}), (\ref{p2gf}), (\ref{p2Hf}), and (\ref{p2H}) (which is a sum
of Eqs.\ (\ref{p2v}) and (\ref{p2u})), the complete NLO contribution to the pressure is
\begin{equation}
p_{2}(T,\mu_q) = - \left[ g^2 (N_c^2-1)N_c + v N_f^2 (N_f^2+1) + 2 u \,N_f^3 \right]\, 
\frac{T^4}{144}
- \left[ g^2 (N_c^2-1)N_f+ y^2 N_f^2 N_c \right]  
\left( \frac{5T^4}{576} + \frac{\mu_q^2 T^2}{32 \pi^2} + \frac{\mu_q^4}{64 \pi^4}\right)\;.
\end{equation}
Using the properly normalised large $N_c$ and $N_f$ couplings  and taking the 
Veneziano limit we arrive at 
\begin{equation}
\frac{p_{2}(T,\mu_q)}{p_{0,g}(T)}= - 5 \left[ \alpha_g  + (\alpha_v  + 2\alpha_h) 
\frac{N_f^2}{N_c^2} \right]\, 
 - 5\left( \alpha_g \frac{N_f}{N_c}+ \alpha_y \frac{N_f^2}{N_c^2}  \right)  
\left( \frac{5}{4} + \frac{9}{2\pi^2}\frac{\mu_q^2}{T^2} 
+ \frac{9}{4\pi^4} \frac{ \mu_q^4}{T^4}\right)\;.
\end{equation}
 Trading $N_f/N_c$ for $\epsilon$ we have: 
\begin{equation}
\frac{p_{2}(T,\mu_q)}{p_{0,g}(T)}= - 5 \left[ \alpha_g  + (\alpha_v  + 2\alpha_h) 
\left(\frac{11}{2} + \epsilon \right)^2 \right]
 - 5\left[ \alpha_g \left( \frac{11}{2} + \epsilon \right)+ \alpha_y 
 \left(  \frac{11}{2}+  \epsilon  \right)^2\right]  
\left( \frac{5}{4} + \frac{9}{2\pi^2}\frac{\mu_q^2}{T^2} 
+ \frac{9}{4\pi^4} \frac{ \mu_q^4}{T^4}\right)\;.
\end{equation}
We observe the explicit dependence on the couplings of the theory. 

 \section{Pressure to NNLO}
\label{ANNLO}
\begin{figure}[h]
\begin{center}
\includegraphics[width=10cm]{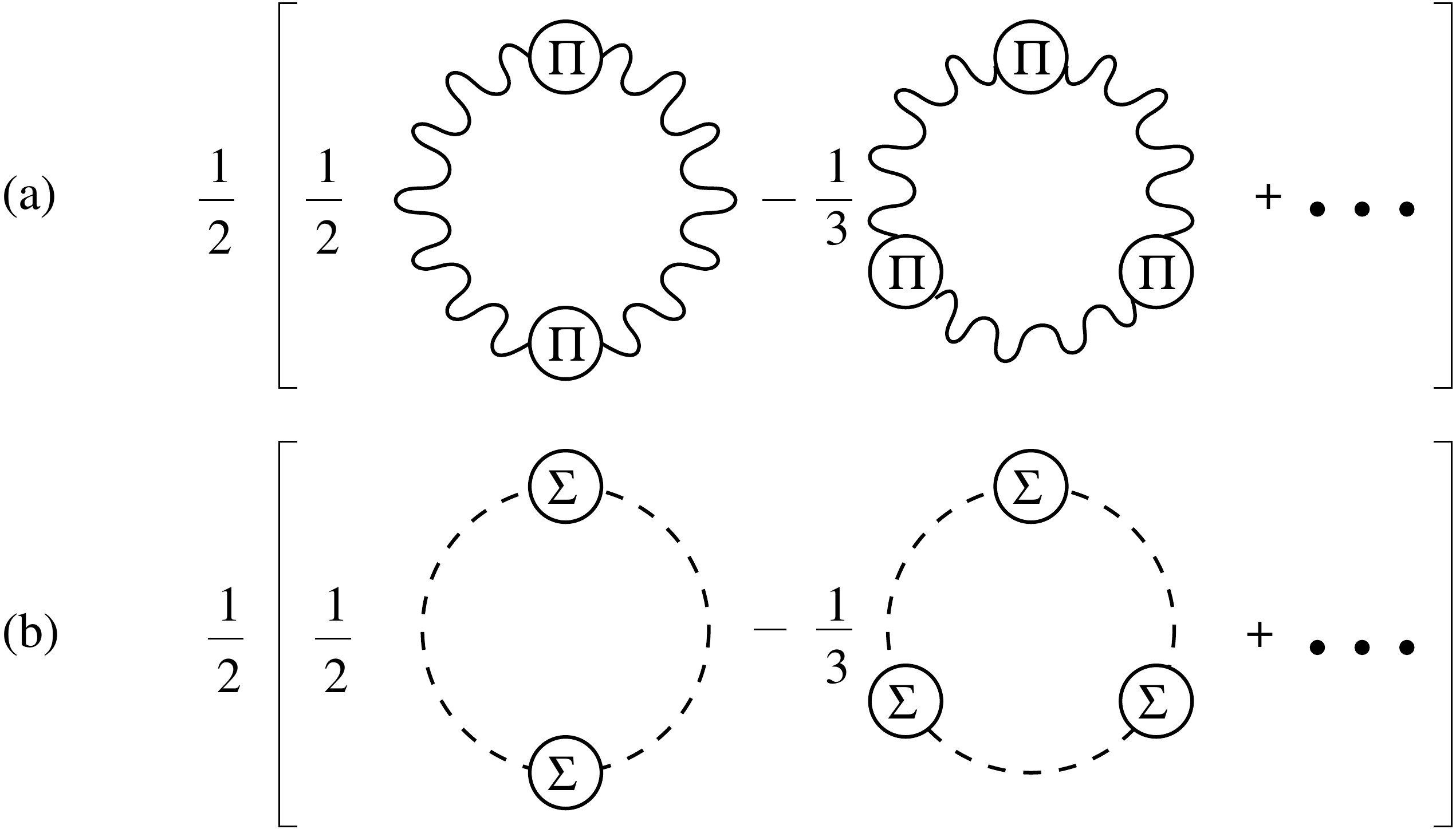}
\caption{Plasmon-ring contributions to the pressure. The diagrams (a) are
the same as in QCD, while the diagrams (b) are the plasmon-ring 
contribution from the scalar field $H$ (shown only for a scalar $S_a$, the one for
a pseudoscalar $P_a$ has the same value).}
\label{fig2}
\end{center}
\end{figure}

To NNLO, the pressure at nonzero temperature and chemical potential
receives contributions from the
so-called plasmon ring diagrams \cite{Kapusta:2006pm}, cf.\ Fig.\ \ref{fig2}.
The leading contribution of these plasmon ring diagrams are $\sim O(g^3)$ and
$\sim O(u^{3/2}, v^{3/2})$. These are calculated in the following.
There is a plasmon ring for the gauge field $A_\mu^i$ and one for the scalar field $H$,
\begin{equation}
p_{3}(T,\mu_q) = p_{3,g}(T,\mu_q) + p_{3,H}(T,\mu_q)\;.
\end{equation}
The contribution from the gauge field, Fig.\ \ref{fig2}(a), is the same as in QCD
\cite{Kapusta:2006pm}, thus
\begin{equation} \label{p3g}
p_{3,g}(T,\mu_q) = (N_c^2-1)\, \frac{T m_g^3(T,\mu_q)}{12 \pi}\;,
\end{equation}
with the electric screening mass of the gauge field \cite{Kapusta:2006pm}
\begin{equation}
m_g^2(T,\mu_q) =g^2 \left[ N_c \frac{T^2}{3} + 
N_f \left( \frac{T^2}{6} + \frac{\mu_q^2}{2\pi^2} \right)\right]\;.
\end{equation}
In analogy to Eq.\ (\ref{p3g}), the plasmon-ring contribution from the scalar field $H$ reads 
\begin{equation}
p_{3,H}(T,\mu_q) = 2 N_f^2\, \frac{T m_H^3(T,\mu_q)}{12 \pi}\;,
\end{equation}
where
\begin{equation}
m_H^2(T,\mu_q) \equiv \Sigma(0,0)
\end{equation}
is the screening mass of the scalar field; $\Sigma(0,0)$ is the zero-Matsubara frequency,
zero-momentum limit of the corresponding one-loop self-energy. The prefactor
$2 N_f^2$ takes into account that we have $2 N_f^2$ (pseudo-)scalar degrees of freedom.
Thus, we only need to compute $\Sigma(0,0)$ for one of these fields, say the scalar
field $S_0$. In general, the one-loop self-energy of the scalar field can be computed by
functional differentiation of the two-loop contribution to the pressure with respect to
the scalar propagator ${\cal S}$ \cite{Kapusta:2006pm},
\begin{equation}
\Sigma \sim - 2\, \frac{\delta p_2}{\delta {\cal S}}\;.
\end{equation}
This corresponds to amputating a scalar propagator in the diagrams shown in Fig.\ \ref{fig1}.
Obviously, only the diagrams (e), (g), and (h) can contribute to $\Sigma$. 
The contribution from the fermion loop in Fig.\ \ref{fig1}(e) is
\begin{eqnarray}
\Sigma_{f}(Q) & = & y^2 T \sum_n \int \frac{{\rm d}^3\vec{k}}{(2 \pi)^3}
\, {\rm Tr} \left[ {\cal G}(K) T_0 {\cal G}(K-Q) T_0 \right] \nonumber \\
& = & \frac{y^2 N_c}{2} \int \frac{{\rm d}^3\vec{k}}{(2 \pi)^3}\,\left\{
(1 - \hat{k} \cdot \hat{p}) \left[ \frac{n_F(k-\mu_q) - n_F(p-\mu_q)}{q_0 + k - p} - 
\frac{n_F(k+\mu_q) - n_F(p+\mu_q)}{q_0 - k + p}  \right] \right. \nonumber \\
 &   & \hspace*{2.3cm} \left. - (1 + \hat{k} \cdot \hat{p}) 
 \left[ \frac{1- n_F(k-\mu_q) - n_F(p+\mu_q)}{q_0 + k + p} - 
\frac{1-n_F(k+\mu_q) - n_F(p-\mu_q)}{q_0 - k - p}  \right] \right\}\;.
\end{eqnarray}
Here, $\vec{p} \equiv \vec{k} - \vec{q}$, $\hat{k} = \vec{k}/k$, and $n_F(k\mp \mu_q)
= [e^{(k\mp\mu_q)/T}+1]^{-1} $ is the Fermi-Dirac distribution for (anti-)particles.
After renormalization of the vacuum contribution and taking the limit $q_0 = 0, \,\vec{q}
\rightarrow 0$, we obtain
\begin{equation} \label{Sf}
\Sigma_f(0,0) = \frac{y^2 N_c}{4} \left( \frac{T^2}{3} + \frac{\mu_q^2}{\pi^2} \right)\;. 
\end{equation}
The contributions from the double-bubble diagrams, Figs.\ \ref{fig1}(g) and (h), can be
decomposed into two parts, one proportional to the vertex $v$ and one to the vertex $u$.
For the first one, we have from the Hartree-type routing of internal indices in Fig.\ \ref{fig1}(g)
a factor $2 \times N_f^2$ (a factor 2 because one can open either one of the two tadpoles), 
and from the Fock-type routing a factor $2\times2=4$ 
(one factor of 2 because the Fock-type diagram appears twice relative to the 
Hartree-type one and one factor of 2 because one can open either one of the two tadpoles).
Finally, we have a factor $2 \times 1 =2$ from the diagram in Fig.\ \ref{fig1}(h)
(one factor of 2 because this type of diagram appears twice relative to Hartree-type one, and
a factor of 1 because functional differentiation only opens the scalar tadpole). 
The remaining tadpole is the same, since
there is nothing that distinguishes scalar from pseudoscalar fields. After 
renormalization of the tadpole (which then is equal to $T^2/12$) we obtain
\begin{equation} \label{Sv}
\Sigma_v = - 2 \left(-\frac{v}{4} \right) \left( 2 \, N_f^2 + 4  +2 \right) \frac{T^2}{12}
= v (N_f^2+3) \, \frac{T^2}{12}\;.
\end{equation}
Finally, we compute the contribution from the diagrams in Figs.\ \ref{fig1} (g) and (h)
proportional to the vertex $u$. Here we go back to the left-hand side of 
Eq.\ (\ref{20}). Amputating a scalar propagator (corresponding to the zeroth
scalar field $S_0$) means that any two of the indices $a,b,c,d$ 
may take the value 0. For symmetry reasons we may restrict ourselves to the
first factor $\delta_{ab} \delta_{cd}$ in the second set of parentheses and multiply
the result by a factor of 3. Now we have either $a=b=0$ or $c=d=0$. This gives
\begin{equation}
\frac{1}{8} \left( d_{abn} d_{cdn} + d_{acn} d_{bdn} + d_{adn} d_{bcn} \right)
\left(\delta_{a0}\delta_{b0} \delta_{cd} + \delta_{c0} \delta_{d0} \delta_{ab} \right)
= \frac{1}{4} \left(d_{aan} d_{00n} + 2 d_{a0n} d_{a0n} \right)
= \frac{1}{2} \left(N_f + \frac{2}{N_f} N_f^2 \right) = \frac{3}{2}\, N_f\;.
\end{equation}
We also need to consider Eq.\ (\ref{21}). Here, only the indices $a,b$ can take the
value 0, and thus we have
\begin{equation}
\frac{1}{4} \left( d_{abn} d_{cdn} + f_{acn} f_{bdn} + f_{adn} f_{bcn} \right) \delta_{a0}
\delta_{b0} \delta_{cd} = \frac{1}{4}\, d_{00n} d_{ccn} = \frac{1}{2}\, N_f\;. 
\end{equation}
Putting everything together, we have
\begin{equation} \label{Su}
\Sigma_u = 2u\, N_f\, \frac{T^2}{12}\;.
\end{equation}
Adding Eqs.\ (\ref{Sf}), (\ref{Sv}), and (\ref{Su}), we finally get
\begin{equation}
m_H^2 (T,\mu_q) \equiv \left[ v (N_f^2+3) + 2 u\, N_f \right] \frac{T^2}{12} + 
\frac{y^2 N_c}{4} \left( \frac{T^2}{3} + \frac{\mu_q^2}{\pi^2} \right)\;. 
\end{equation}

\end{document}